\documentclass[journal=jacsat,manuscript=article]{achemso}

\usepackage[version=3]{mhchem} 
\usepackage{graphicx}
\usepackage{dcolumn}
\usepackage{bm}
\usepackage{xcolor}
\usepackage{pdfpages}

\author{Da Wang}
\affiliation{Department of Physics, Columbia University, New York, NY 10027, USA}
\alsoaffiliation{Department of Mechanical Engineering, Columbia University, New York, NY 10027, USA}
\altaffiliation{These authors contributed equally to this work}

\author{Evan J. Telford}
\affiliation{Department of Physics, Columbia University, New York, NY 10027, USA}
\altaffiliation{These authors contributed equally to this work}

\author{Avishai Benyamini}
\affiliation{Department of Physics, Columbia University, New York, NY 10027, USA}
\alsoaffiliation{Department of Mechanical Engineering, Columbia University, New York, NY 10027, USA}

\author{John Jesudasan}
\affiliation{Tata Institute of Fundamental Research, Homi Bhabha Road, Colaba, Mumbai 400 005 India}

\author{Pratap Raychaudhuri}
\affiliation{Tata Institute of Fundamental Research, Homi Bhabha Road, Colaba, Mumbai 400 005 India}

\author{Kenji Watanabe}
\affiliation{National Institute for Materials Science, 1-1 Namiki, Tsukuba, 305-0044 Japan}

\author{Takashi Taniguchi}
\affiliation{National Institute for Materials Science, 1-1 Namiki, Tsukuba, 305-0044 Japan}

\author{James Hone}
\affiliation{Department of Mechanical Engineering, Columbia University, New York, NY 10027, USA}

\author{Cory R. Dean}
\affiliation{Department of Physics, Columbia University, New York, NY 10027, USA}
\email{cd2478@columbia.edu}

\author{Abhay N. Pasupathy}
\affiliation{Department of Physics, Columbia University, New York, NY 10027, USA}
\alsoaffiliation{Condensed Matter Physics and Materials Science Department, Brookhaven National Laboratory, Upton, NY 11973, USA}
\email{apn2108@columbia.edu}
\title{{Andreev Reflections in NbN/graphene Junctions under Large Magnetic Fields}}

\abbreviations{SC, superconductor; g, graphene; QH, quantum Hall;  NbN, Niobium Nitride; SAR, specular Andreev reflection; RAR, retro Andreev reflection; FQHE, fractional quantum Hall effect; hBN, hexagonal Boron Nitride; CNP, charge neutrality point; e-beam, electron beam; BTK, Blonder-Tinkham-Klapwijk; BdG, Bogoliubov-de-Gennes}

\begin{document}
\begin{abstract}
  Hybrid superconductor/graphene (SC/g) junctions are excellent candidates for investigating correlations between Cooper pairs and quantum Hall (QH) edge modes. Experimental studies are challenging as Andreev reflections are extremely sensitive to junction disorder and high magnetic fields are required to form QH edge states. We fabricated {{low-resistance}} SC/g interfaces, composed of graphene edge contacted with NbN with a barrier strength of $Z\approx 0.4$, that remain superconducting under magnetic fields larger than $18$ T. We establish the role of graphene's Dirac band structure on zero-field Andreev reflections and demonstrate dynamic tunability of the Andreev reflection spectrum by moving the boundary between specular and retro Andreev reflections with parallel magnetic fields. Through the application of perpendicular magnetic fields, we observe an oscillatory suppression of the 2-probe conductance in the $\nu = 4$ Landau level attributed to the reduced efficiency of Andreev processes at the {{NbN/g}} interface, consistent with theoretical predictions.
\end{abstract}

\section{Keywords}
two-dimensional, van der Waals, graphene, niobium nitride, superconductivity, Andreev reflection, quantum Hall effect, Zeeman splitting

\section{Text}
At the interface between a normal metal and a superconductor, electrons in the normal metal with energy below the superconducting gap can only transition into the superconductor via Andreev reflection, where an electron incident on the interface reflects as a hole along with the transmission of a Cooper pair into the superconductor. The probability of this process strongly depends on the band structure of the normal metal and the properties of the interface. Graphene, due to its Dirac band structure with a highly tunable Fermi level, is an enticing avenue to explore the role of the normal metal's electronic properties on Andreev processes\cite{Beenakker2006, Efetov2016theory, Efetov2016exp, Soori2018, Sahu2016, Sahu2018}. Initial experiments by Efetov in 2016 demonstrated the effect of graphene's Dirac electronic structure on Andreev processes, identifying intraband Andreev processes (RAR) and interband Andreev processes (SAR). Under the application of perpendicular magnetic fields, graphene hosts chiral QH edge states, which has allowed for the investigation of Andreev process from integer QH states\cite{Amet2016,Calado2015,Gunel2014, Sahu2018}, which manifest non-Abelian zero modes of Majorana fermions\cite{Lutchyn2010,Sarma2015,San-Jose2015,Fu2008,Sahu2018}. However, there are still many unanswered questions such as how junction transparency and the proximity effect in graphene evolve under large perpendicular magnetic fields in the integer and fractional QH regime and the role of interfacial vortices on the junction properties. In addition the effect of large parallel magnetic fields on the Andreev reflection spectrum (where Zeeman splitting is induced without the formation of QH states) has thus far been unexplored. The limitations in experiments on SC/g junctions arise from technical challenges in fabricating SC/g junctions with favorable properties. {{Previous reports on SC/g junctions suffer either from relatively low junction transparency (as Andreev reflections are extremely sensitive to interfacial inhomogeneity and disorder\cite{Blonder1982,Daghero2010}) or low critical magnetic fields ($\leq8$ T), limited by the superconducting contact.\cite{Efetov2016exp,Amet2016}}}

\begin{figure*}
    \centering
    \includegraphics[width=1.0\textwidth]{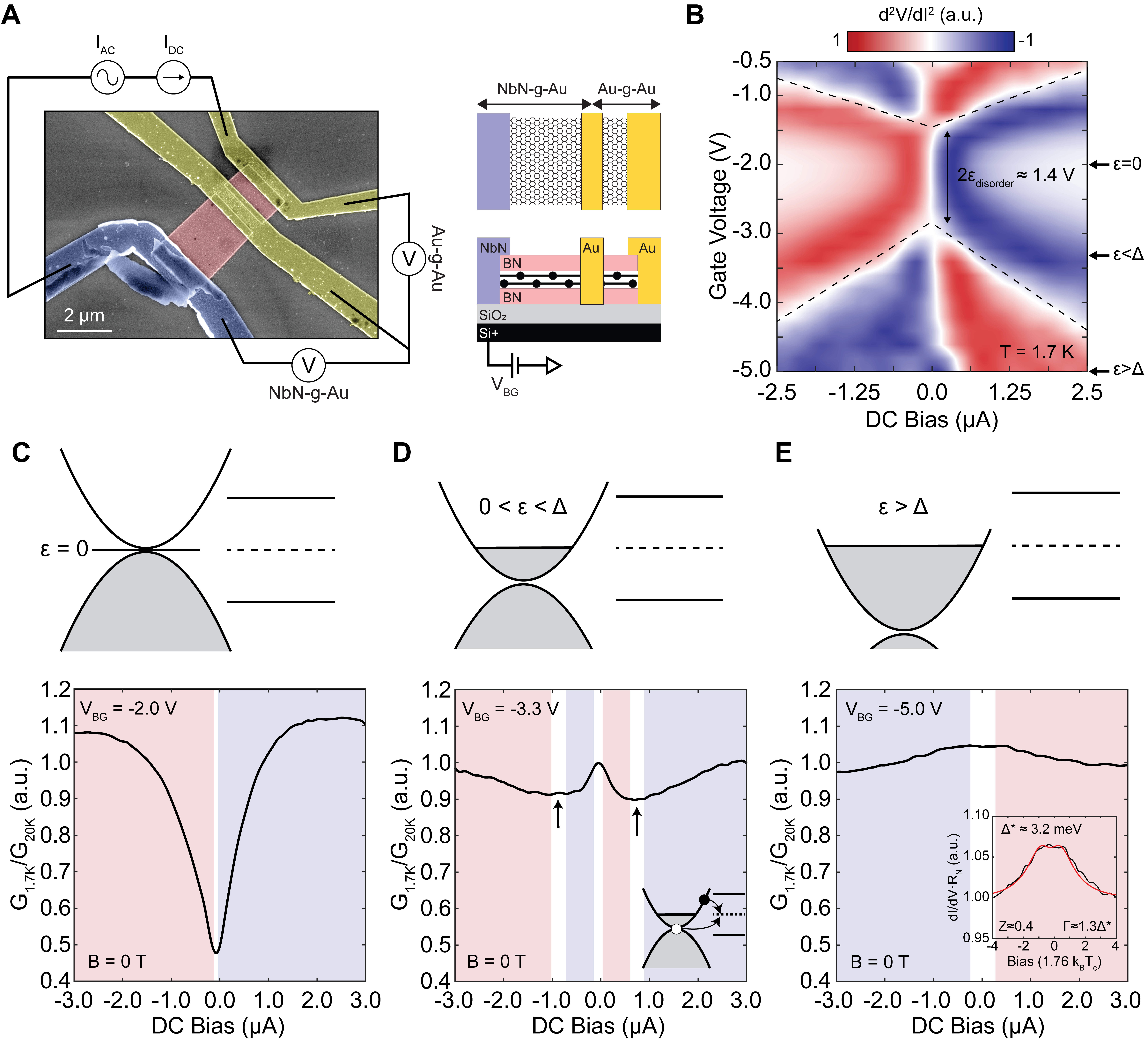}
    \caption{\small{Andreev reflection in a NbN/g junction at zero magnetic field. (A) {{A scanning electron microscopy (SEM) image of the measured device with the corresponding measurement configuration (left) and a top and side view schematic of the device geometry (right). Color code: blue, NbN leads; yellow, Au leads; red, hBN; grey, SiO\textsubscript{2}; black, Si.}} (B) Derivative of differential junction resistance versus DC current bias and back-gate voltage. The dashed black lines denote the crossover from SAR to RAR regimes. Arrows label the position in gate voltage of the cuts in C-E. (C-E) differential conductance versus DC current bias for $V_{BG}  = -2.0$ V (C), $V_{BG}=-3.3$ V (D), and $V_{BG}=-5.0$ V (E). Blue, red, and white colored regions correspond to increasing, decreasing, and constant differential conductance, respectively, in panel (B). A cartoon of the band structure at the {{NbN/g}} interface is given above each plot. The left side of each cartoon is an energy versus momentum diagram of the graphene band structure, while the right side is an energy versus real space diagram of the NbN band structure. The inset of (D) shows a cartoon of the Andreev processes when $\epsilon_F = e|V_{NS}|$. {{The injected electron (black dot) and Andreev reflected hole (white dot) have an energy of $\epsilon_{e/h} = \epsilon_F \pm eV_{NS}$.}} The inset of (E) plots the normalized junction conductance (solid black line) versus DC voltage bias with a fit to the BTK model (solid red line). Extracted junction parameters are given in the inset. The extracted SC gap of NbN is larger than expected{{ ($\Delta^* > \Delta = 1.76\cdot k_BT_C \approx 2$ meV)}} as the voltage bias is not entirely dropped across the NbN/g junction.}}
    \label{fig: 1}
\end{figure*}

In this paper, we report SC/g junctions fabricated using Niobium Nitride (NbN) as an edge contact to bilayer graphene fully encapsulated with hexagonal Boron Nitride (hBN)\cite{Dean2010,Wang2013}. Bilayer graphene is used due to the smaller disorder broadening of the charge neutrality point (CNP) compared to monolayer graphene\cite{Efetov2016theory,Efetov2016exp,Soori2018}. Figure \ref{fig: 1}A shows the schematic of the device. First we use the dry-polymer-transfer technique to make a fully encapsulated graphene heterostructure\cite{Wang2013}, followed by electron beam (e-beam) lithography and reactive ion etching to define the device area. Superconducting ($2$ nm Ti + $100$ nm NbN) and normal metal ($2$ nm Cr + $90$ nm Au) edge contacts are patterned by e-beam lithography and deposited by e-beam deposition (Cr/Ti/Au) and sputtering (NbN)\cite{PratapSputtering}. By sweeping the back-gate voltage and total current bias, we have full control of the Fermi energy in graphene and voltage bias across the NbN/g junction. NbN has a large, isotropic upper critical field\cite{Mondal2011} ({{figure}} S1), which allows for the investigation of finite-field phenomena such as vortex formation at the NbN/g edge, junction transparency as a function of magnetic field, and the role of Zeeman splitting on Andreev reflections. By investigating the perpendicular magnetic field dependence of the junction resistance, we observe a suppression of the conductance within the QH plateaus, which can be well explained by interfacial scattering at a non-ideal SC/2DEG interface using the Blonder-Tinkham-Klapwijk (BTK) theory to model the scattering of QH edge states\cite{Hoppe2000,Eroms2005}. Studying the junction under parallel magnetic fields, we observe shifts in the Andreev reflection spectrum in graphene, which is explained by Zeeman splitting of the graphene band structure.

We begin by investigating Andreev reflections at the {{NbN/g}} interface through non-equilibrium conductivity measurements at zero magnetic field in figure \ref{fig: 1} (see supplemental for measurement details). By varying the electronic density in the graphene through electrostatic gating, we can access three distinct tunneling regimes (demonstrated in figures \ref{fig: 1}C-E). In regime 1, the Fermi level of graphene is at the CNP (cartoon in figure \ref{fig: 1}C). The {{NbN/g}} junction conductance manifests a local minimum at zero bias due to the vanishing graphene density of state (DOS). As the DC current bias is increased, the conductance increases as charge carriers are injected at a finite energy above or below the CNP (figure \ref{fig: 1}C{{; temperature dependence in figure S2}}). In regime 2, the Fermi level of the graphene is tuned away from the CNP, but within the superconducting gap of the NbN (cartoon in figure \ref{fig: 1}D). We observe a peak in the differential conductance at zero bias which is attributed to Andreev processes as the feature is suppressed for $T>T_C$ ({{figure S3}}). We also observe local differential conductance minima when the Fermi level of graphene is equal to the voltage bias across the junction interface ($\epsilon_{F}=e|V_{NS}|$) (figure \ref{fig: 1}D)\cite{Efetov2016theory,Efetov2016exp}. In regime 3, the Fermi level of graphene is further tuned away from the CNP and outside of the superconducting gap of NbN. We observe a zero-bias differential conductance peak due to Andreev reflections. We no longer observe minima in the differential conductance as the CNP is outside of the superconducting gap and the graphene behaves as a normal metal. In this regime, we can quantify the transparency of the junction using BTK theory (see supplemental for details)\cite{Blonder1982}. We find the barrier strength of our junction to be $Z\approx 0.4$ with a broadening parameter of {{$\Gamma \approx 1.3\Delta^*$}}, comparable to other {{reported SC/g junctions}}\cite{Efetov2016theory,Amet2016,Lee2017}. The large value of the broadening parameter $\Gamma$ may be due to NbN/g interface disorder created during fabrication, doping inhomogeneities, or external electromagnetic noise. 

Figure \ref{fig: 1}B shows a two-dimensional (2D) colormap of the derivative of the differential resistance $\left (\frac{d^2 V}{dI^2}\right )$ versus DC bias and back-gate voltage. This emphasizes the boundaries between the three observed tunneling regimes. The dashed black lines in figure \ref{fig: 1}B clearly define the boundary between SAR and RAR as previously observed in NbSe\textsubscript{2}/graphene junctions\cite{Efetov2016theory,Efetov2016exp} {{(see figure S4 for temperature dependence)}}. In the absence of disorder, the boundary between Andreev reflection regimes should intersect at $\epsilon_F = e|V_{DC}| = 0$. In our junction, we observe a finite width at $V_{DC}=0$ due to disorder in the graphene, which we estimate to be $V_{disorder}\approx 0.7$ V, similar to previously reported NbSe\textsubscript{2}/graphene junctions\cite{Efetov2016exp}.

\begin{figure*}
    \centering
    \includegraphics[width=1.0\textwidth]{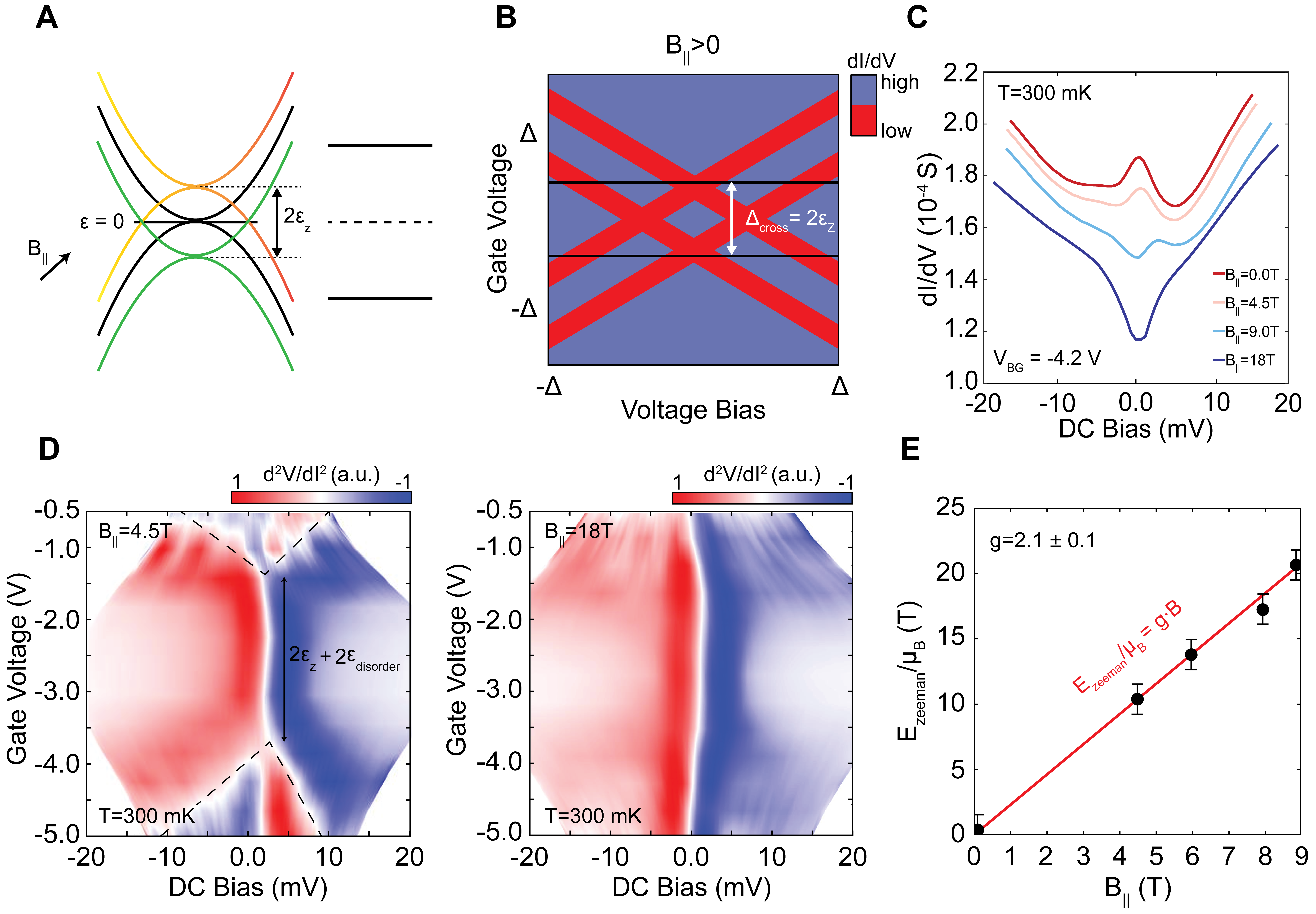}
    \caption{\small{Andreev reflection in a NbN/g junction under parallel magnetic field. (A) Schematic diagram of energy vs. momentum at the NbN/g interface under an in-plane magnetic field. Zeeman splitting is denoted as $\epsilon_Z$. Unsplit bands are shown as solid black lines while spin up (down) bands are shown in orange (green). (B) Schematic of the anticipated differential conductance versus DC bias and back-gate voltage under an applied in-plane magnetic field. (C) Differential conductance versus DC voltage bias with $V_{BG}=-4.2$ V for various in-plane magnetic fields. (D) Colormap of the derivative of junction resistance $\left ( \frac{d^2V}{dI^2} \right )$ versus DC voltage bias and back-gate voltage at an in-plane magnetic field of $4.5$ T (left) and $18$ T (right). DC voltage bias was converted from DC current bias by integrating the measured differential resistance and using $V=I\cdot R$. The dashed black lines denote the boundary between SAR and RAR regimes. (E) Zeeman splitting extracted by measuring the boundary between SAR and RAR (dashed black line in (D)) versus in-plane magnetic field. The extracted $g$ factor is $g=2.1\pm 0.1$.}}
    \label{fig: 2}
\end{figure*}

We next explore the effect of parallel magnetic field as it is predicted that Zeeman splitting of graphene's Dirac cones will lead to the observation of specular Andreev processes at zero DC bias\cite{Soori2018}. Figure \ref{fig: 2}A shows a schematic energy vs. momentum diagram of the NbN/g interface under an in-plane magnetic field. Zeeman splitting breaks the spin degeneracy, energetically separating the graphene bands into two copies with opposite spins (see {{figure S5}} for details). The boundary defining the crossover between RAR and SAR processes shifts correspondingly (figure \ref{fig: 2}B). A regime of SAR (diamond-shaped blue region in figure \ref{fig: 2}B) appears at zero bias. In that regime, injected electrons with spin up in the valence band are reflected as holes with spin down in the conduction band\cite{Lutchyn2010}. To illustrate the movement of the boundary between SAR and RAR, cuts of differential conductance versus DC bias at selected magnetic fields are shown in figure \ref{fig: 2}C. We choose a fixed back-gate voltage of $V_{BG} = -4.2$ V, where we clearly observe the SAR/RAR boundary at zero magnetic field (see {{figure S6}} for cuts at additional back-gate values). At a parallel field of $4.5$ T, the conductance enhancement due to Andreev reflections at zero bias decreases. {{When the parallel field is increased further to $9$ T, the zero-bias conductance peak has transitioned into a conductance dip, indicating a transition from tunneling regime 2 (figure 1D) to tunneling regime 1 (figure 1C). At higher fields up to $18$ T, only a conductance dip is observed near zero bias likely becuase the NbN superconducting gap is below the graphene Fermi level (figure S8).}} By plotting the derivative of the differential resistance $\left (\frac{d^2 V}{dI^2}\right )$ versus DC bias and back-gate voltage (as in figure \ref{fig: 1}B) at different in-plane magnetic fields, one can extrapolate how the boundary between tunneling regime 1 and regime 2 moves as a function of in-plane field (figure \ref{fig: 2}D and {{figure S7}}). We define the boundary between the two tunneling regimes as the $V_{BG}$ at which a zero-bias conductance peak appears. With no disorder, one would expect the zero-bias conductance peaks to appear for any finite gate bias away from the CNP (with $\epsilon_{F}<\Delta$). The role of disorder is to broaden the CNP so the zero-bias conductance peaks appear at finite gate biases away from the CNP. We assume the broadening is independent of magnetic field, so any changes in the position of the boundary relative to the zero-field position is attributed to Zeeman splitting. In figure \ref{fig: 2}D, the boundary has moved from $V_{BG} = -3.4$ V (at zero field, {{figure S7}}) to $V_{BG} = -3.8$ V at $4.5$ T (figure \ref{fig: 2}D left) and disappears completely at $18$ T (figure \ref{fig: 2}D right). By tracking the position of the boundary for different in-plane magnetic fields, we extract the induced Zeeman splitting ({{figure \ref{fig: 2}E}}). The induced Zeeman splitting is linear with the applied field and we extract an effective $g$ factor of $g=2.1\pm 0.1$.

\begin{figure*}
    \centering
    \includegraphics[width=1.0\textwidth]{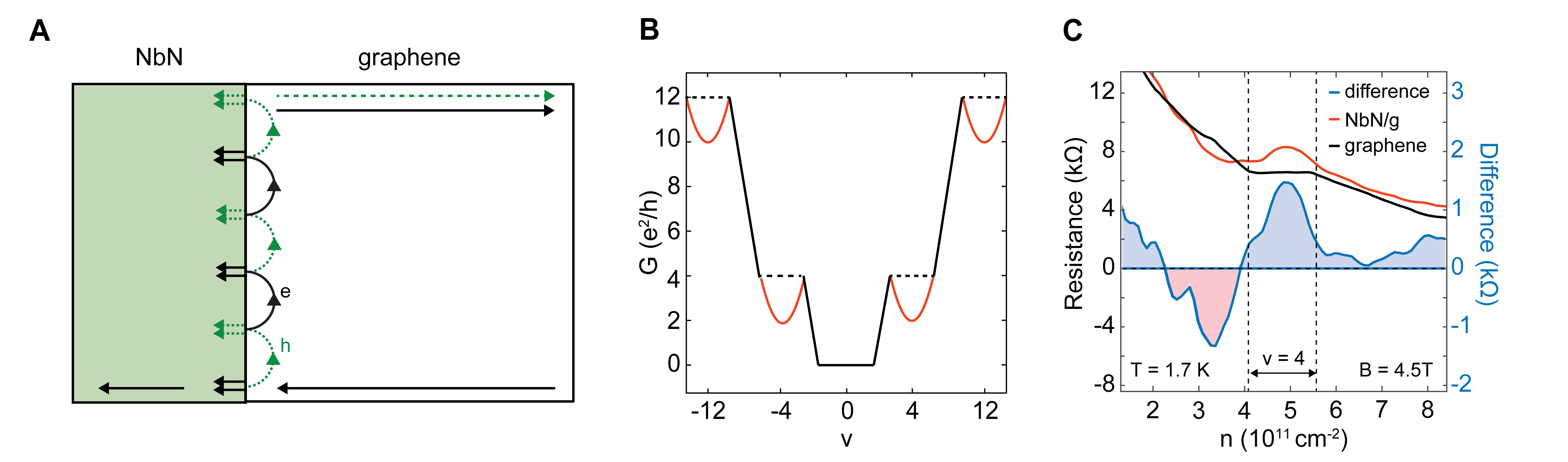}
    \caption{\small{Intersection of the QHE and superconductivity. (A) A schematic of the classical skipping orbit picture at the NbN/g interface. Electron (hole) trajectories are given by solid black (green dashed) lines. NbN and graphene are denoted by green and grey regions, respectively. (B) Schematic of conductance versus filling factor for a SC/g junction in the QH regime with a perfect interface transparency $w=0$ (solid and dashed black line) and a non-ideal interface transparency $w>0$ (solid black line and solid red line)\cite{Hoppe2000}. (C) graphene channel (solid black curve) and NbN/g (solid red curve) junction resistance versus electron density measured at a perpendicular magnetic field of $4.5$ T. In each curve, the contact resistance was extracted at zero field at a graphene carrier density of $n\approx 2.4\times 10^{12}$ cm\textsuperscript{-2} and subtracted from the presented data. The difference between the two curves is given by the solid blue curve. Highlighted blue and red regions signify when the NbN/g junction resistance is larger and smaller than the graphene channel (Au/g) resistance, respectively. The vertical dashed black lines demarcate the $\nu=4$ plateau.}}
    \label{fig: 3}
\end{figure*}

By applying perpendicular magnetic fields, we can investigate correlations between superconductivity and the QHE. In the QH regime, the bulk conductance of graphene vanishes, and electrons are only transported through edge conductance channels. A semi-classical skipping orbit picture can be used to understand the Andreev process that {{occur}} at the {{NbN/g}} interface (figure \ref{fig: 3}A). Electrons are injected at one edge of the interface and reflected as holes (green) as a result of intraband RAR process. Through a similar Andreev process, the reflected holes are re-reflected as electrons at the {{NbN/g}} interface. For long junction widths ($W\gg R_C$, where $W$ is the junction width and $R_C$ is the cyclotron radius), an equilibrium superposition of electrons and holes is formed at the SC/g interface known as an Andreev bound state. As Sch\"{o}n et al. pointed out, the Büttiker description of quantum transport can be used to calculate the Andreev contribution to the conductance at a SC/2DEG interface in the QH regime.

\begin{equation}
G_{AR}=  \frac{e^2}{\pi \hbar} \sum_1^{n^*} B_n 
\label{equ: 1}
\end{equation}
$B_n$ is the hole probability for a particular Andreev bound state and the summation is over the Andreev bound states that intersect with the chemical potential\cite{Hoppe2000}. The energy spectrum of Andreev bound states (and hole probabilities $B_n$) at the SC/2DEG interface is found by solving the Bogoliubov-de-Gennes (BdG) equation with spatially non-uniform single-electron/hole Hamiltonians and pair potential\cite{Hoppe2000}. For an ideal interface without scattering, QH plateaus are predicted (figure \ref{fig: 3}B). For a non-ideal interface, we expect an oscillation of the conductance within a given Landau level (figure \ref{fig: 3}B). 

To properly quantify the role of Andreev processes in the QH regime, we compare measurements of the NbN/g and graphene channel resistances versus Landau level filling factor for the same device (figure \ref{fig: 3}C). Contact resistances of $\approx 246 \Omega$ for the NbN/g junction and $\approx 104 \Omega$ for the Au/g junction were measured at a graphene carrier density of $n\approx 2.4\times 10^{12} $ cm\textsuperscript{-2} at zero magnetic field and zero DC bias. The contact resistances increase with magnetic field, but the difference in contact resistance between Au/g and {{NbN/g}} is nearly constant versus field ({{figure S9}}). Outside of the QH plateaus, the NbN/g and graphene channel display a nearly constant difference related to the discrepancy in contact resistance between Au and NbN for a given magnetic field. When graphene is tuned to the $\nu = 4$ QH state, the difference in conductance between graphene channel and NbN/g channel is prominent and depends upon carrier density in an oscillatory fashion. We can understand these observations by noting that Andreev processes are sensitive to the band structure of graphene\cite{Hoppe2000}. Outside of the QH plateaus, conductance is dominated by bulk transport in which normal Andreev processes occur, whereas inside the QH plateaus, the only available Andreev process is through the Andreev bound edge states (figure \ref{fig: 3}A). 

\begin{figure*}
    \centering
    \includegraphics[width=1.0\textwidth]{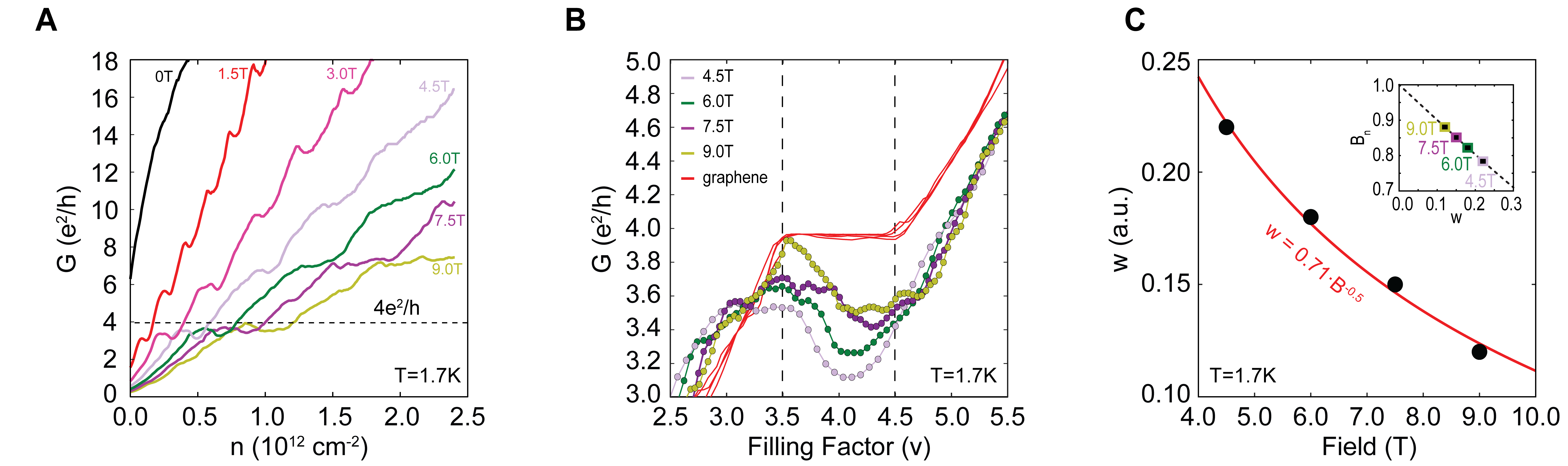}
    \caption{\small{NbN/g junction transparency vs. perpendicular magnetic field. (A) NbN/g junction conductance versus graphene carrier density for various perpendicular magnetic field values. The $\nu = 4$ plateau is marked by a black dashed line. (B) NbN/g (yellow, purple, pink, and green solid lines) and graphene channel (red lines) conductance versus graphene filling factor at various perpendicular magnetic fields. The $\nu=4$ plateau is marked by the black dashed lines. (C) Extracted scattering parameter $w$ versus perpendicular magnetic field. The solid red line is a fit to the theoretically predicated $w\propto \frac{1}{\sqrt{B}}$ dependence. Inset: Extracted hole probability amplitude versus scattering parameter $w$ for the data in (B). The dashed black line is a fit to the data with equation \ref{equ: 2}}.}
    \label{fig: 4}
\end{figure*}

In figure \ref{fig: 4}, we characterize the quality of the NbN/g junction by examining the perpendicular magnetic field dependence of the junction conductance. In figure \ref{fig: 4}A, we plot the NbN/g junction conductance versus graphene carrier density at various perpendicular magnetic field values. The QH plateaus are not perfectly quantized (compared to our graphene channel in {{figure S10}}), which indicates a finite junction transparency. For analysis of the NbN/g junction quality, we focus on the $\nu = 4$  plateau as it is well developed in the graphene channel for all presented magnetic field values (figure \ref{fig: 4}B). From Sch\"{o}n et.al, the two important parameters used to describe the quality of the junction are the Fermi velocity mismatch $s=\sqrt{\frac{\epsilon_F^{S} m_N}{\epsilon_F^N m_S}}$ (where $\epsilon_F^S$, $\epsilon_F^N$ and $m_S$, $m_N$ are the Fermi energies and effective electron masses for the superconductor and normal metal respectively) and the interfacial barrier strength $w=\sqrt{\frac{2m_N U_0^2}{\hbar^2\epsilon_F^N}}$ (where $U_0$ is the interface scattering potential), which quantifies the interface scattering. For an ideal interface without backscattering and no Fermi velocity mismatch, $s=1$, $w=0$.  For simplicity, we set $s=1$ since it is magnetic field independent and our primary objective is to determine the magnetic field dependence of $w$. The barrier strength $w$ can be determined by extracting the minimal NbN/g junction conductance within the $\nu = 4$  plateau region as the junction conductance depends only on $w$ and $\nu$.
\begin{equation}
    G_{AR}=\frac{e^2}{\pi \hbar} \sum_1^4 B_n = 2 \frac{e^2}{\pi \hbar} \frac{q^2/(1- \gamma_0^2)}{(1+ \sqrt{1-q^2/(1- \gamma_0^2)})}
    \label{equ: 2}
\end{equation} 
\begin{equation}
    q=\frac{2s}{s^2+w^2+1}
    \label{equ: 3}
\end{equation}
\begin{equation}
    \gamma_0=\frac{(s^2+w^2-1) \sin(\pi \nu/2)+2w \cos(\pi \nu/2) }{s^2+w^2+1}
    \label{equ: 4}
\end{equation}
The scattering parameter $w$ depends upon magnetic field through $w=\sqrt{\frac{2m_N U_0^2}{\hbar^2 \epsilon_F^N (B)}}$, where $\epsilon_F^N (B)= \nu \frac{\hbar \omega_C}{2}$and $\omega_C = \frac{eB}{mc}$ ({{figure S11}}). The inset of figure \ref{fig: 4}C shows extracted hole probability $B_n$ versus scattering parameter $w$. We find $w$ equals $0.22$, $0.18$, $0.15$, $0.12$ for perpendicular field values of $B = 4.5$ T, $6.0$ T, $7.5$ T, $9.0$ T, respectively. We can demonstrate the data is well described by Sch\"{o}n’s model\cite{Hoppe2000} by fitting the extracted $w$ versus magnetic field and comparing it to the expected $w\propto \frac{1}{\sqrt{B}}$ dependence. We find close agreement between our experiment and theory (figure \ref{fig: 4}C).

Although Sch\"{o}n’s model\cite{Hoppe2000} fits our data, it omits the contribution of higher order corrections to the conductance due to phase coherence in wide SC/2DEG junctions (defined as $W\gg R_C,$ where $W$ is the junction width and $R_C$ is the cyclotron radius). It is worth discussing the role of phase coherence and why our data can be accurately described by the first order coherence term. When examining the SC/2DEG interface, the four tunneling processes we consider are e$\to$e, e$\to$h, h$\to$e, h$\to$h, whose corresponding probabilities are ($r_{ee}$, $r_{eh}$, $r_{he}$, $r_{hh}$) . To take phase coherence into consideration, the transmission matrix  is generalized to include the spatial dependence of the superconductor phase $\phi(y)$ and to consider all possible scattering trajectories (in the form of individual action terms for electrons $S_e$ and holes $S_h$)\cite{Chtchelkatchev2007}. The final expression for junction conductance can be written as 
\begin{equation}
    G(\nu) = \sum_{n=0}^{\infty} g_n \cos{\left (2\pi \nu n + \delta_n\right )}
    \label{equ: 5}
\end{equation} 
where $\delta_n$ is related to the phase coherence. The higher order harmonics ($n$\textsuperscript{th} term corresponding to $n$ coherent e/h bounces) are relevant for larger junction widths but are suppressed by the presence of interfacial disorder. The leading term to the conductance can be written:
\begin{equation}
    G(\nu)\approx g_0 + g_1 \cos{\left (2\pi \nu +\delta_1 \right )}
    \label{equ: 6}
\end{equation}
An intuitive way to understand the suppression of higher order terms by interface disorder is the following: at the junction, short range disorder is considered by introducing fluctuations that re-scatter electrons and holes, destroying phase coherence between multiple bounces. Therefore the conductance oscillations are predominantly described by the first order harmonic (approximation used in Sch\"{o}n's model\cite{Hoppe2000}) when the average length scale of the interface disorder and inhomogeneity is less than the cyclotron radius $R_C=hk_{F}/2\pi eB$. 

In conclusion, {{low-resistance}} junctions between NbN, a superconductor with an isotropic critical magnetic field exceeding $18$ T, and bilayer graphene are fabricated and investigated. At zero field, we observe the role of graphene’s Dirac cones on Andreev reflections, demonstrating 3 distinct tunneling regimes. With the application of parallel fields, we demonstrate dynamic tunability of the boundary between RAR and SAR. The movement of the boundary corresponds to the expected Zeeman splitting in bilayer graphene, from which we extract a $g$ factor of $g\approx 2.1 \pm 0.1$. Finally, through the application of perpendicular fields, we observe an oscillatory suppression of the 2-probe conductance in the $\nu=4$ Landau level that is well described by a theoretical model given by Sch\"{o}n et al\cite{Hoppe2000}. For future experiments, realizing NbN/g junctions with lower graphene disorder broadening will allow us to better explore the regime of zero-field SAR near the CNP and investigate the intersection of superconductivity and the FQHE. Furthermore, if $\nu = 0$ helical edge modes in graphene are stabilized\cite{Maher2013,Young2014}, counter propagating edge modes of quantum spin hall states can be realized at the NbN/g interface, which is predicted to host non-abelian Majorana physics\cite{Young2014,Hart2014,Qi2011,Wiedenmann2016,Bocquillon2017}.

\section{Supporting Information}
The following files are available free of charge.
\\
Fabrication Methods, Transport Measurements, Fitting Zero-Field NbN/g/Au Junction Conductance to BTK Theory, Modeling NbN/g/Au Junction Conductance in the QH Regime, {{Supporting Figures S1-S12 (PDF)}}

\section{Corresponding Author}
*Abhay N. Pasupathy (e-mail: apn2108@columbia.edu) \\
*Cory R. Dean (e-mail: cd2478@columbia.edu)

\section{Notes}
the authors declare no competing financial interests.

\section{Author Contributions}
The manuscript was written through contributions of all authors. All authors have given approval to the final version of the manuscript. \\
DW and EJT contributed equally
\section{Funding Sources}

This research was supported by the Columbia MRSEC on Precision-Assembled Quantum Materials (PAQM) - DMR-2011738, Honda Research Institute USA Inc and the Air Force Office of Scientific Research via grant FA9550-21-1-0378. A portion of this work was performed at the National High Magnetic Field Laboratory, which is supported by National Science Foundation Cooperative Agreement No. DMR-1157490 and the State of Florida. This work was performed in part at the Advanced Science Research Center NanoFabrication Facility of the Graduate Center at the City University of New York. The authors from TIFR would like to thank  Department of Atomic Energy, Government of India (Grant No. 12-R\&D-TFR-5.10-0100) for financial support.

\section{Abbreviations}
SC, superconductor; g, graphene; QH, quantum Hall;  NbN, Niobium Nitride; SAR, specular Andreev reflection; RAR, retro Andreev reflection; FQHE, fractional quantum Hall effect; hBN, hexagonal Boron Nitride; CNP, charge neutrality point; e-beam, electron beam; BTK, Blonder-Tinkham-Klapwijk; DOS, density of states; 2D, two-dimensional; DC, direct current; AC, alternating current; QHE, quantum Hall effect; 2DEG, two-dimensional electron gas; BdG, Bogoliubov-de-Gennes
\newpage

\bibliography{nl-2021-01072m_Manuscript_Final_Revisions_Editor_Comments}


\begin{tocentry}

 \includegraphics[width=1.0\textwidth]{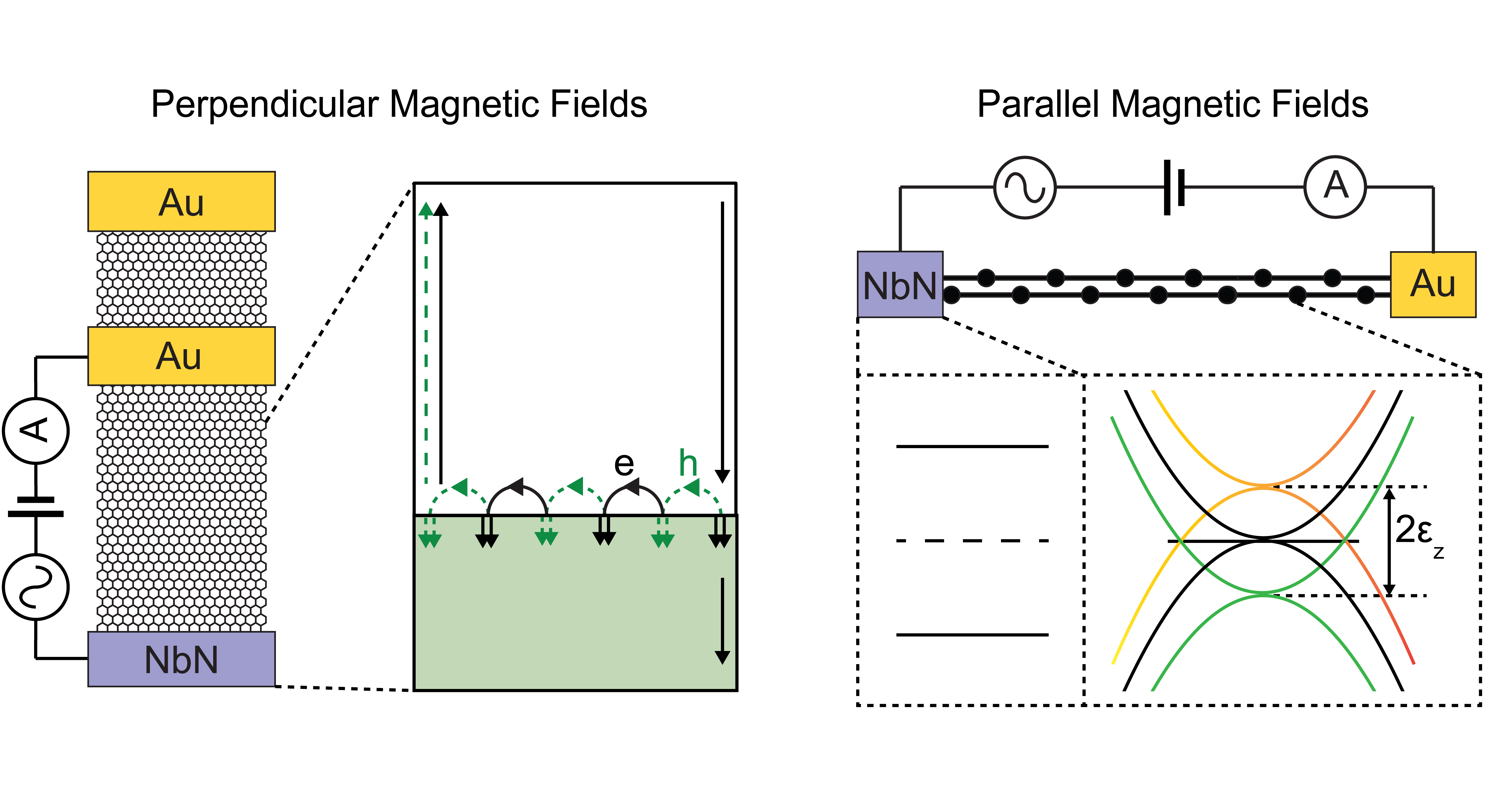}

\end{tocentry}

\includepdf[pages=-]{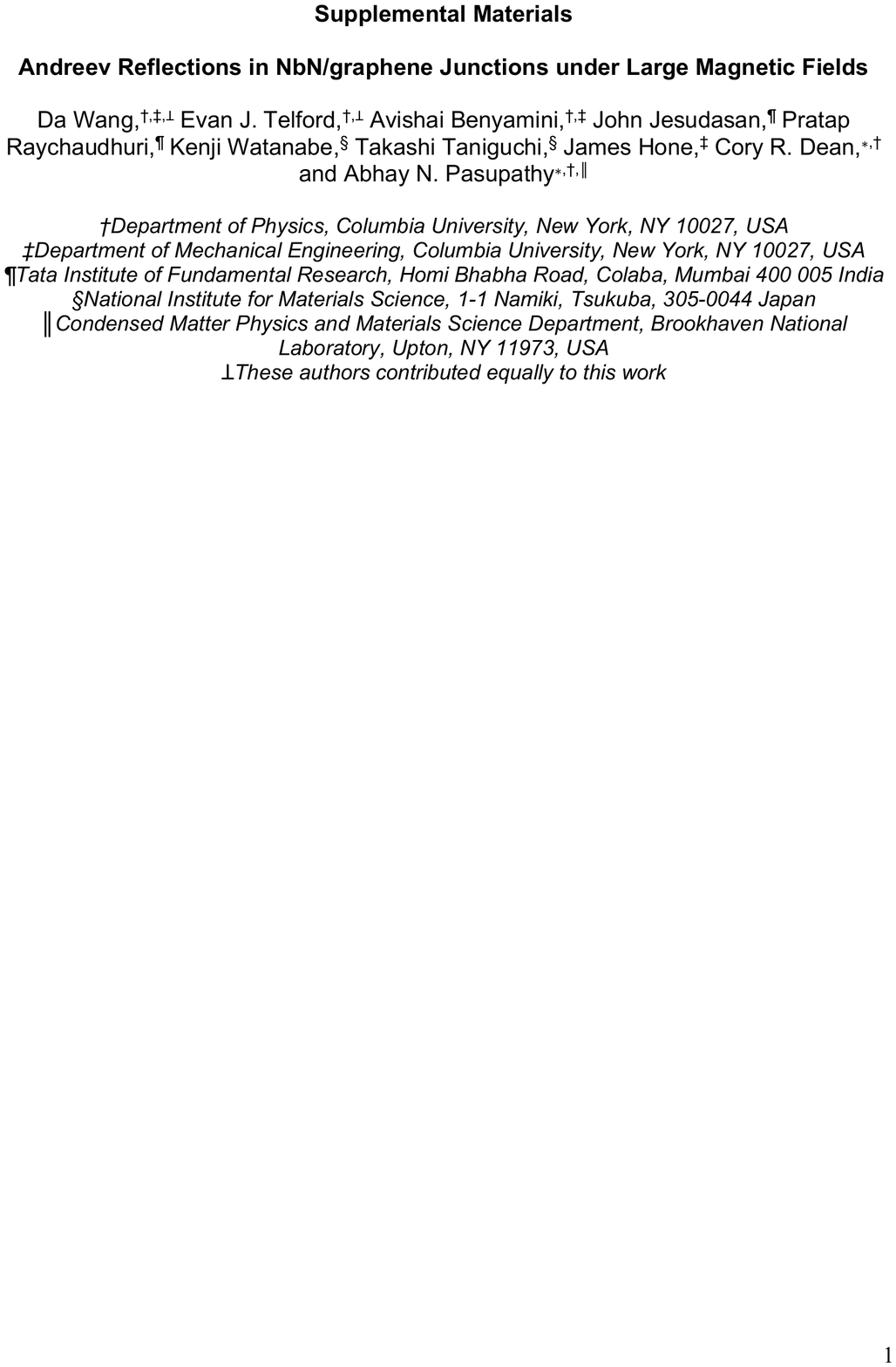}

\end{document}